\documentclass[a4paper,twocolumn,11pt]{quantumarticle}
\pdfoutput=1
\usepackage[utf8]{inputenc}
\usepackage[english]{babel}
\usepackage[T1]{fontenc}
\usepackage{amsmath}
\usepackage{hyperref}

\usepackage{tikz}
\usepackage{lipsum}

\usepackage{cite}
\usepackage{braket}
\usepackage{graphicx}
\usepackage{mathtools}
\usepackage{listings}
\usepackage[square,numbers,sort&compress]{natbib}

\graphicspath{ {./images/} }
\DeclarePairedDelimiter\floor{\lfloor}{\rfloor}
\begin{document}

\title{Applying Grover's Algorithm to Hash Functions: \newline A Software Perspective}

\author{Richard Preston}
\affiliation{© 2022 The MITRE Corporation, Bedford, MA 01730, USA. All Rights Reserved. Approved for Public Release; Distribution Unlimited. Public Release Case \#22-0014}
\orcid{0000-0002-1137-082X}

\maketitle

\begin{abstract}
  Quantum software frameworks provide software engineers with the tools to study quantum algorithms as applied to practical problems. We implement classical hash functions MD5, SHA-1, SHA-2, and SHA-3 as quantum oracles to study the computational resource requirements of conducting a preimage attack with Grover's Algorithm. We introduce an improvement to the SHA-3 oracle that reduces the number of logical qubits required in the Keccak block permutation by 40\%.
\end{abstract}

\section{Introduction} \label{introduction}

Consider the following scenario. On a certain banking website, customers log in by entering their username and password. Before being sent to the server for authentication, the password is salted (random data is appended) and then passed through a cryptographic hash function.\footnote{A hash function is an algorithm designed to scramble data in a deterministic fashion, so that (1) the same input always produces the same output, but (2) given the output, it is difficult to determine what input produces it. This is also called a “one-way” or “trapdoor” function because it is relatively easy to compute in one direction, but practically impossible to go backwards.}

Suppose an attacker gains access to the plaintext hash and salt. With conventional (classical) computers, the only way to recover the original password is with guess-and-check, a.k.a.\ brute-force. Assuming an 8-character password consisting of random characters, the attacker would have to check approximately $2^{64}=1.8\times10^{19}$ combinations. At a rate of 1 billion checks per second, it would take on-average 292 years to find the password, at which point it is probably useless.

But what if the attacker had access to a powerful quantum computer? With Grover's Algorithm, the password could be found directly in $\sqrt{2^{64}}=4.3\times10^9$ iterations~\cite{10.1145/237814.237866}. Assuming the same rate of 1 billion iterations per second, this could be accomplished in a matter of seconds. The attacker could then log in and steal the customer's funds before the breach is detected. Fortunately in this scenario, today's quantum computers are not capable of performing computations of this scale. However, it is critically important to understand precisely what computational resources \textit{would} be required for such an attack.

The objective of this research was to study the computational cost of conducting a preimage attack on MD5, SHA-1, SHA-2, and SHA-3 with Grover's Algorithm. We used Microsoft's Quantum Development Kit (QDK) to implement each hash function as a quantum oracle, validate the implementations through simulation, and then estimate the quantum hardware requirements to apply Grover's Algorithm.

\subsection{Background} \label{background}

Grover's Algorithm can reverse a black-box function implemented as a quantum oracle in $\mathrm{O}(\sqrt{N})$ iterations with $\mathrm{O}(\log_2{N})$ qubits, with $N$ being the number of possible input combinations to the function~\cite{10.1145/237814.237866}. In this context, the quantum oracle phase-flips a target qubit when the desired output is produced (e.g., when the password is correct). Grover's Algorithm searches for the input(s) that cause the phase-flip to occur.

The steps of the algorithm are given below. Assume $n$ is the number of input bits, $N=2^n$, and $k$ is the number of ``correct'' input combinations or search targets.

\begin{enumerate}
  \item Allocate $n$ input qubits and 1 target qubit.
  $$\ket{\mathrm{I},\mathrm{T}} = \ket{0^{\otimes n},0}$$
  \item Apply H to each input qubit.
  $$\ket{\mathrm{I},\mathrm{T}} = \frac{1}{\sqrt{N}}\sum_{i=0}^{N-1}\ket{i} \otimes \ket{0}$$
  \item Apply X to the target qubit.
  $$\ket{\mathrm{I},\mathrm{T}} = \frac{1}{\sqrt{N}}\sum_{i=0}^{N-1}\ket{i} \otimes \ket{1}$$
  \item Do the following $\floor*{\frac{\pi}{4}\sqrt{\frac{N}{k}}}$ times:
  \begin{enumerate}
    \item Apply the oracle.
    \item Apply H to each input qubit.
    \item Apply Z to the target qubit, zero-controlled on all input qubits.
    \item Apply H to each input qubit.
  \end{enumerate}
  \item Measure the input qubits. The result will be a value that produces the desired output with probability approaching 1 for large $N$. Check the outcome and repeat if the search failed.
\end{enumerate}

\begin{figure}[htbp]
  \centering
  \includegraphics[width=\linewidth]{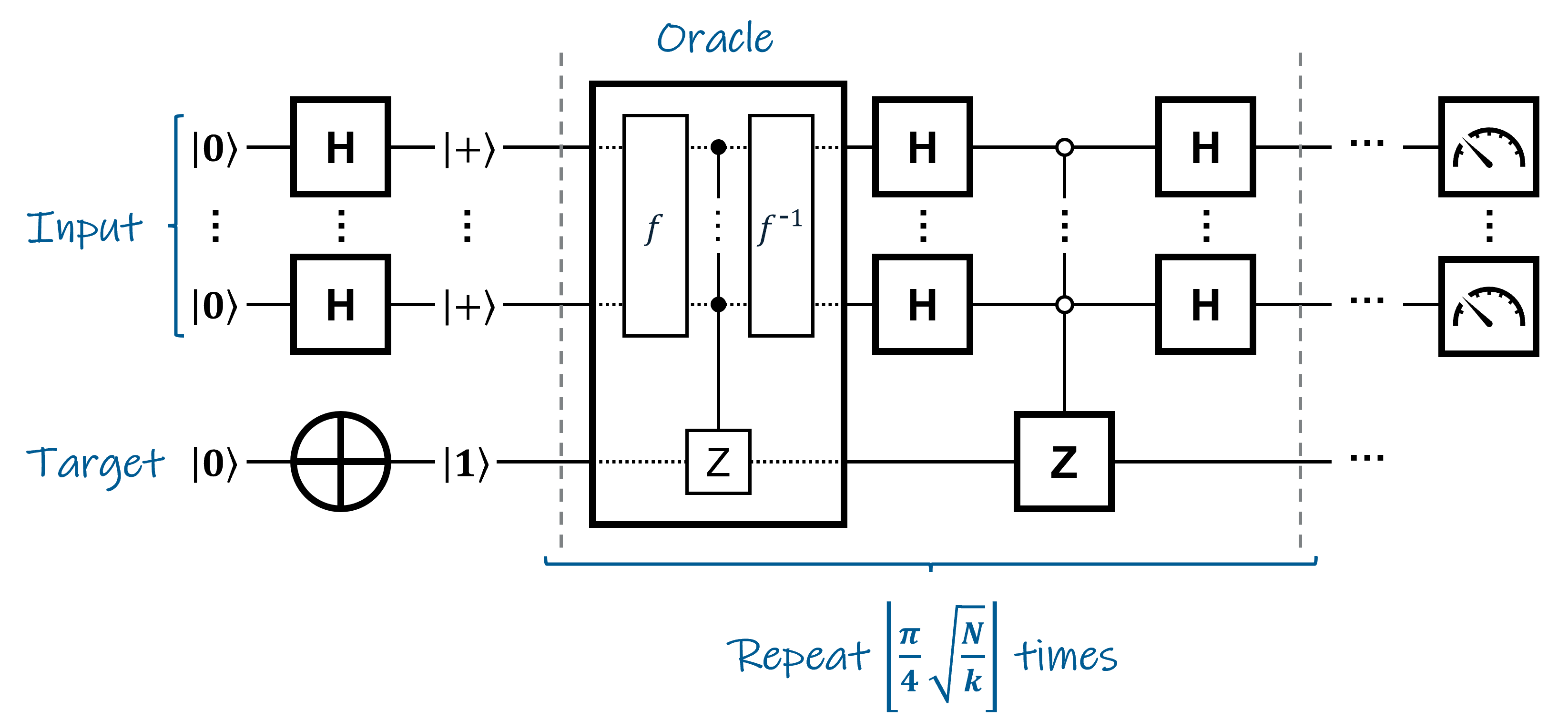}
  \caption{Grover's Algorithm}
  \label{fig:grovers}
\end{figure}

Fig.~\ref{fig:grovers} shows the circuit diagram for Grover's Algorithm. Note how the oracle is constructed such that any operations applied in computing the black-box function are reversed.

The challenge in applying Grover's Algorithm to a hash function lies in translating the classical algorithm into a quantum one. Referring again to Fig.~\ref{fig:grovers}, the $f$ and $f^{-1}$ in the oracle are not given – they must be designed using quantum computing primitives. Section~\ref{approach} details how MD5, SHA-1, SHA-2, and SHA-3 can be translated into quantum oracles so the quantum preimage attack can be conducted.

\subsection{Related Work} \label{related}

The most relevant paper to this work is ``Estimating the cost of generic quantum pre-image attacks on SHA-2 and SHA-3'' (2016), where Amy et al.\ analyze the computational resources required to run Grover's Algorithm on SHA-256 and SHA3-256, assuming a surface code based quantum computer~\cite{amy2016estimating}. They provide oracle implementations for the hash functions and make precise, educated estimates on the number of logical qubits and surface code cycles that would be needed. There are three important ways in which our work differs from theirs:

\begin{enumerate}
  \item Our analysis includes the entire SHA-2 and SHA-3 families, as well as MD5 and SHA-1.
  \item We fully implement the preimage attack as a quantum software program and use a resource estimator to generate the cost metrics. This approach has benefits and limitations, as discussed in Section~\ref{conclusion}.
  \item We introduce an improvement to the SHA-3 implementation that reduces the number of logical qubits required.
\end{enumerate}

Our research is an example of a ``practicality assessment'' of a quantum computing application. Rather than introduce a novel quantum algorithm, a practicality assessment seeks to precisely understand the computational requirements of applying an existing algorithm, with all the implementation details laid bare. A similar approach is taken by Joe Clapis in his report, ``A Quantum Dot Plot Generation Algorithm for Pairwise Sequence Alignment''~\cite{clapis2021quantum}. In that work, he finds that the Quantum Pairwise Sequence Alignment algorithm proposed by Prousalis and Konfaos~\cite{Prousalis2019-mr} relies on a black-box operation that overwhelms the rest of the algorithm in terms of computational complexity when a practical implementation is attempted. Such insights are critical as quantum computers become advanced enough to solve real-world problems.

\section{Approach} \label{approach}

The entirety of this work was conducted using Microsoft's QDK, with the majority of the code written in Q\#, Microsoft's programming language for quantum computing~\cite{Svore_2018}. The QDK provides built-in simulation and estimation capabilities that allowed us to validate the correctness of the program before obtaining resource requirement metrics. Of particular use was the Toffoli simulator, which can simulate large numbers of qubits but only allows X and controlled X gates. (It essentially treats the qubits as classical bits.) This allowed us to fully test any operation that was derived from a classical function.

The procedure we followed consists of three steps:

\begin{enumerate}
  \item Implement the hash function as a quantum operation in Q\#.
  \item Validate the correctness of the program using the Toffoli simulator.
  \item Estimate the computational resources required to conduct a preimage attack with Grover's Algorithm.
\end{enumerate}

The first step is the most challenging, since quantum and classical computing occupy two completely different paradigms. It is oftentimes inefficient or even impossible to translate a classical function directly into a quantum one. For example, since measurement destroys superposition, you cannot ``set'' or ``clear'' a qubit like you can a classical bit. This is particularly relevant in managing working memory because any operations performed on temporary qubits must be reversed before they can be reused; you can't just set them to zero directly. You also cannot create an independent copy of a qubit (no-cloning theorem~\cite{wootters1982single}), something that is trivial and highly useful in classical programming.

The remainder of this section describes each hash function studied and highlights the key differences in their classical and quantum implementations.

\subsection{MD5}

The high-level designs of MD5, SHA-1, and SHA-2 are very similar to each other. They are all built using a Merkle–Damgård construction, and therefore contain the following elements~\cite{10.1007/0-387-34805-0_21}\cite{10.1007/0-387-34805-0_39}:

\begin{itemize}
  \item The input message is padded using a special function that extends its length to be a multiple of a certain number (e.g. 512). A property of the padding function is that it encodes the original input length.
  \item The message is broken up into fixed-length blocks or chunks that are processed one at a time.
  \item The chunks are processed with a one-way compression function that mixes two fixed-length inputs into an output of length equal to one of the inputs. Specifically, each chunk is mixed with the output of processing the previous chunk. (There is an initialization vector to mix with the first chunk.)
\end{itemize}

In the case of MD5, SHA-1, and SHA-2, the compression functions are composed of computational iterations that include nonlinear Boolean functions, arithmetic addition, and bitwise rotation. They all follow the same basic pattern but differ in the specific operations employed. In MD5, there is a 128-bit state that is initialized to a fixed value and updated after each chunk is processed. The compression function applied to the chunks consists of 64 iterations that transform a copy of the 128-bit state broken into four 32-bit working registers, $a$, $b$, $c$, and $d$. After the iterations are complete, the registers are recombined, and the result is added to the original state before proceeding to the next chunk.

\begin{figure}[htbp]
  \centering
  \includegraphics[width=\linewidth]{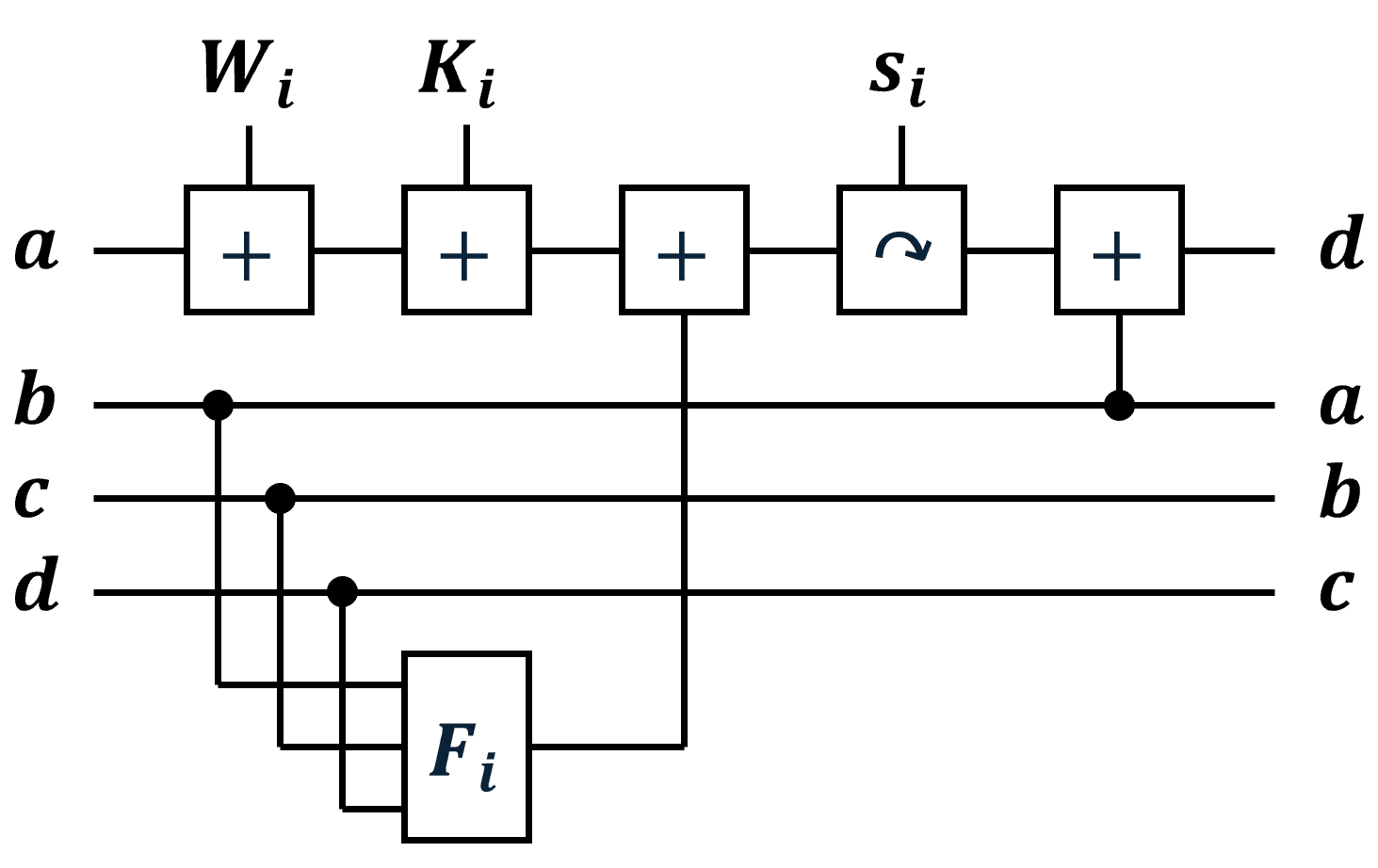}
  \caption{Single MD5 Iteration}
  \label{fig:md5-classical}
\end{figure}

Fig.~\ref{fig:md5-classical} illustrates the computation involved in a single MD5 iteration. In the diagram, the plus sign means arithmetic addition and the clockwise arrow means bitwise left-rotate.\footnote{The result of left-rotating a 32-bit register $r$ by $n$ bits is classically defined as $(r \ll n)\;|\;(r \gg (32 - n))$, where $\ll$ and $\gg$ are bit-shifts and $|$ is bitwise OR.} $W_i$ is a 32-bit word in the message block determined by the iteration index. $K_i$ and $s_i$ are predetermined constants. $F_i$ is one of four nonlinear Boolean functions that are each used for 16 of the 64 iterations per chunk. All values are modulus 32-bit and little-endian. Note how the names of the working registers are shuffled after each iteration. See IETF RFC 1321 for the full MD5 specification~\cite{RFC1321}.

The main considerations when creating a quantum version of a classical algorithm are (1) how the individual operations translate, (2) how to minimize the number of qubits required, and (3) how to ensure that the entire design is reversible, i.e., it supports the adjoint functor in Q\#.\footnote{For more info on functors, see \url{https://docs.microsoft.com/en-us/azure/quantum/user-guide/language/expressions/functorapplication}.} For example, MD5 uses many arithmetic additions. Q\# offers a library function implementing an in-place quantum ripple carry adder, which can be used to handle addition between two qubit registers.\footnote{We used the library function that implements Takahashi et al.'s quantum ripple carry adder design~\cite{takahashi2009quantum}.} However, this cannot be used to add a constant ($K_i$) directly. Instead, it must first be encoded in a temporary qubit register using an in-place XOR, and then added normally. Once the addition is done, the temporary register must be returned to zero by applying the XOR again so the qubits can be de-allocated. This entire process is reversible, so it can be used in the overall design.

\begin{figure}[htbp]
  \centering
  \includegraphics[width=\linewidth]{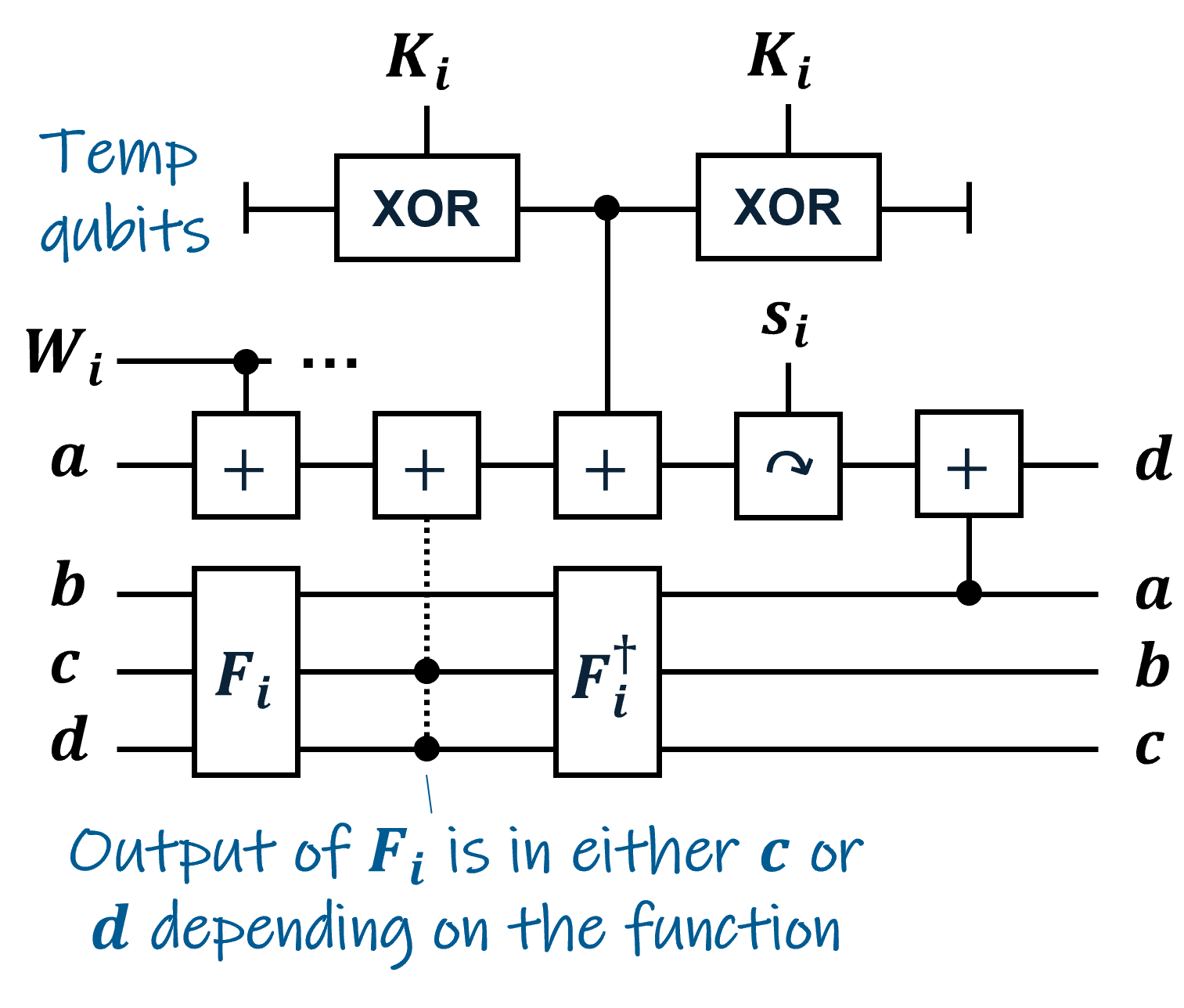}
  \caption{Quantum Version of MD5 Iteration}
  \label{fig:md5-quantum}
\end{figure}

Fig.~\ref{fig:md5-quantum} shows the strategy we employed for implementing MD5 in quantum software. Here, all values are little-endian encoded in 32-qubit registers. The plus sign means in-place arithmetic addition of two qubit registers. The clockwise arrow is a re-indexing of the qubit register which has the effect of a bitwise left-rotate. $W_i$ is a 32-qubit segment of the message block determined by the iteration index. $K_i$ and $s_i$ are the same as before. Quantum versions of $F_i$ perform the Boolean functions in-place, so that the output is stored in $c$ or $d$. $F_i$ is reversed after the result is added to $a$ to preserve the contents of the working registers.

One complication of the quantum implementation of MD5 is that, classically, the working registers are reset after each chunk is processed and their contents are added to the state. This would prevent Grover’s algorithm from being applied because (1) directly resetting a qubit involves measuring it, which would destroy any superposition, and (2) the contents of the working registers are needed for the entire design to be reversible. Thus, 128 workspace qubits are needed per chunk if the adjoint functor is to be supported. This is not so surprising when considering that hash algorithms are designed to be one-way, and so additional state information is needed to reverse them directly. That is, you (obviously) cannot directly reverse a hash algorithm given only the output – otherwise that would defeat the entire purpose!

\subsection{SHA-1}

\begin{figure}[htbp]
  \centering
  \includegraphics[width=\linewidth]{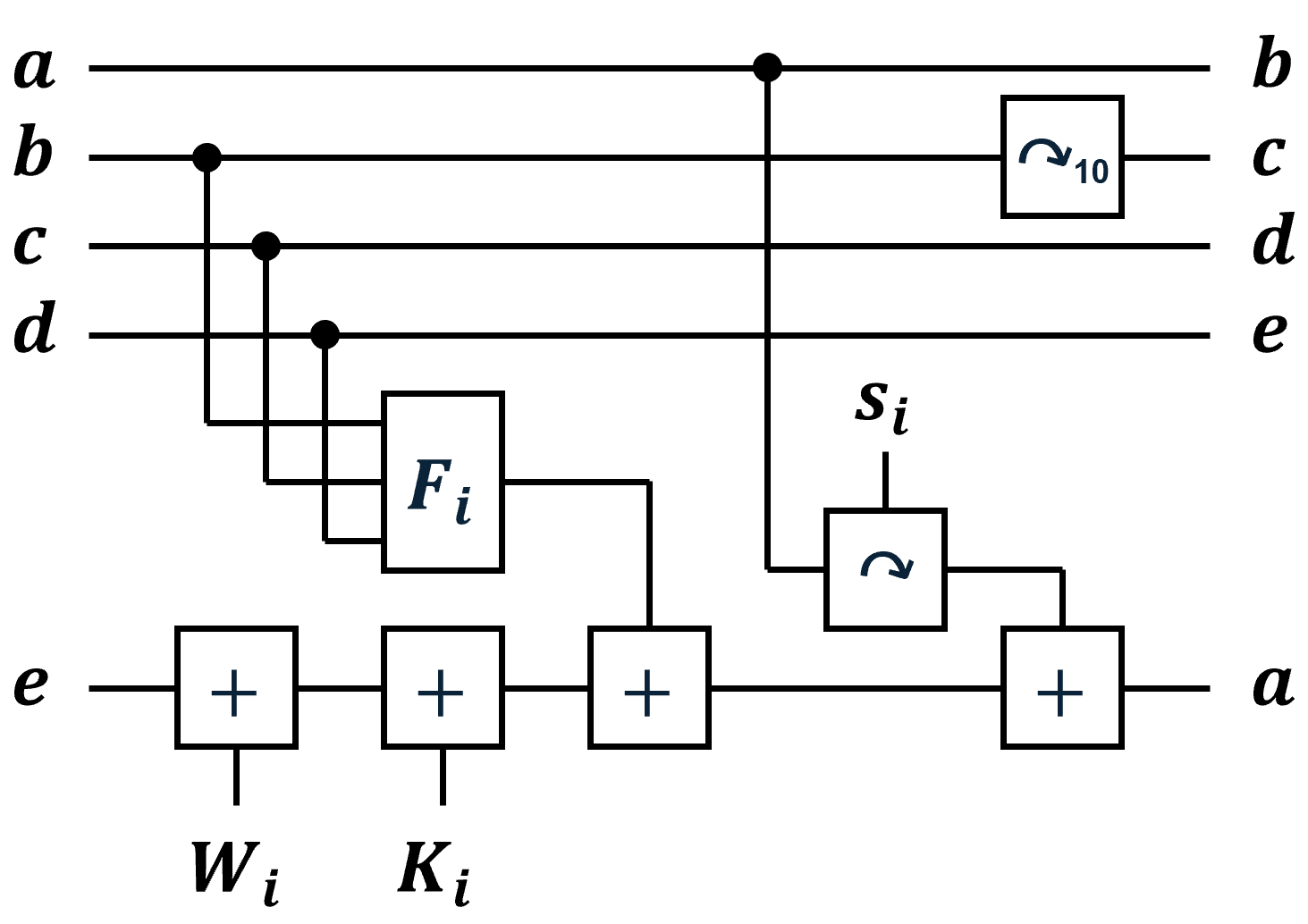}
  \caption{Single SHA-1 Iteration}
  \label{fig:sha1-classical}
\end{figure}

SHA-1 is very similar in structure to MD5, except the state is 160 bits instead of 128, so it uses five 32-bit working registers when processing each chunk. Also, each chunk is extended to 2560 bits (eighty 32-bit words) by mixing the original message block in a deterministic fashion.\footnote{The first 16 words come from the chunk of the padded input message. After that, word $i$ is the bitwise XOR of words $(i-3)$, $(i-8)$, $(i-14)$, and $(i-16)$.} The SHA-1 compression function uses 80 iterations instead of 64, where a single iteration is as illustrated by Fig.~\ref{fig:sha1-classical}. See IETF RFC 3174 for the full specification of SHA-1~\cite{RFC3174}.

The Boolean functions and constants are distinct from MD5, but the general pattern is the same. The working registers are shuffled in a different order, with the additions performed on $e$ instead of $a$. Register $b$ is also bitwise left-rotated by 10 each iteration.

\begin{figure}[htbp]
  \centering
  \includegraphics[width=\linewidth]{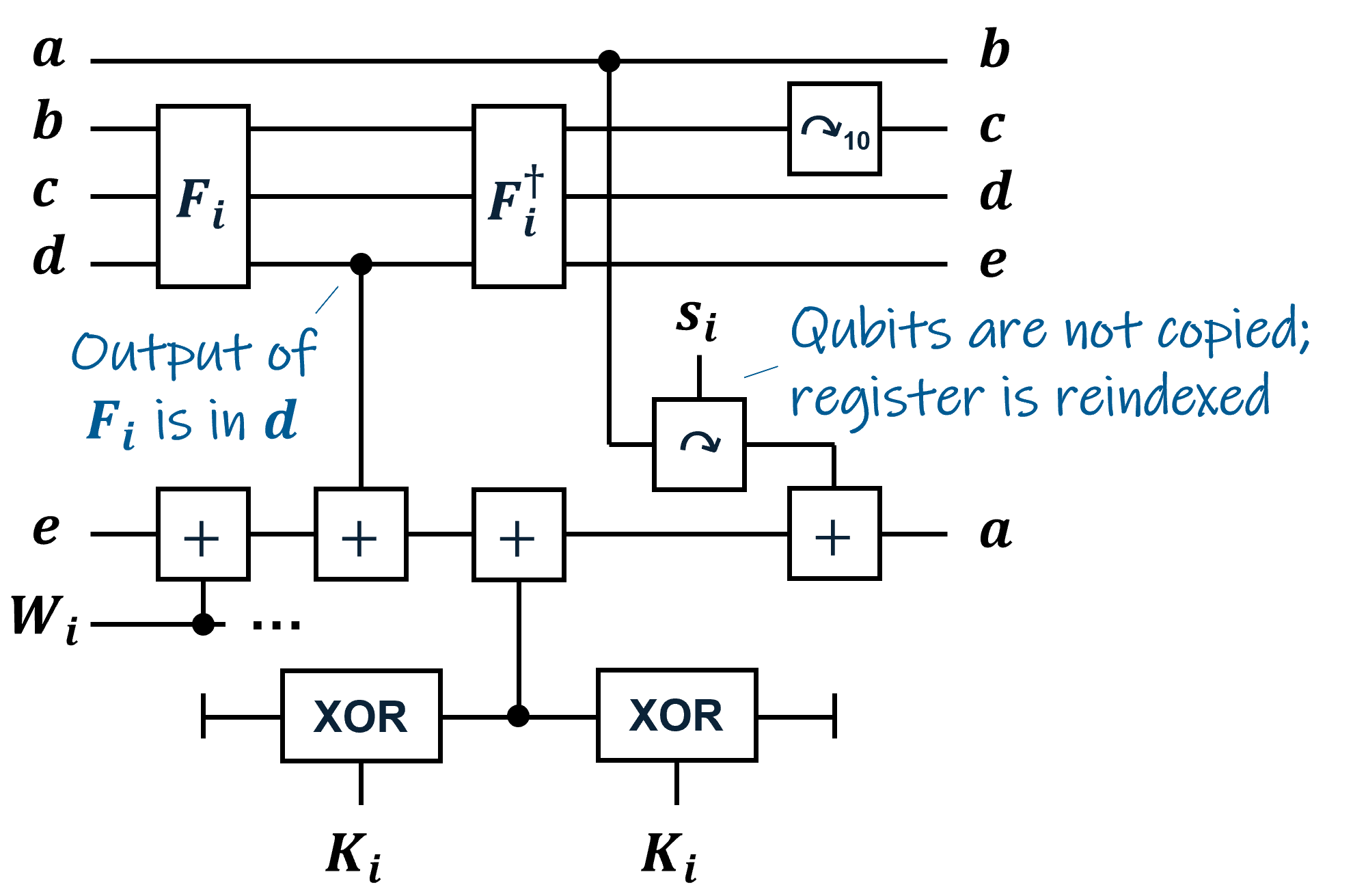}
  \caption{Quantum Version of SHA-1 Iteration}
  \label{fig:sha1-quantum}
\end{figure}

Fig.~\ref{fig:sha1-quantum} shows how we implemented a SHA-1 iteration with similar strategies to MD5. Just as before, each horizontal line represents a little-endian value encoded in a 32-qubit register. (Note that the left-rotated version of $a$ added to $e$ does not count as additional qubits since it is just a re-indexing of the register.) Thus, 160 qubits are required for the working registers, plus 32 temporary qubits to add the constant $K_i$ to $e$. $W_i$ comes from the 80-word message schedule generated from the message block before the iterations begin. We again have the unfortunate situation that the 160 workspace qubits must be preserved after each chunk is processed to make the entire design reversible.

\subsection{SHA-2}

\begin{figure}[htbp]
  \centering
  \includegraphics[width=\linewidth]{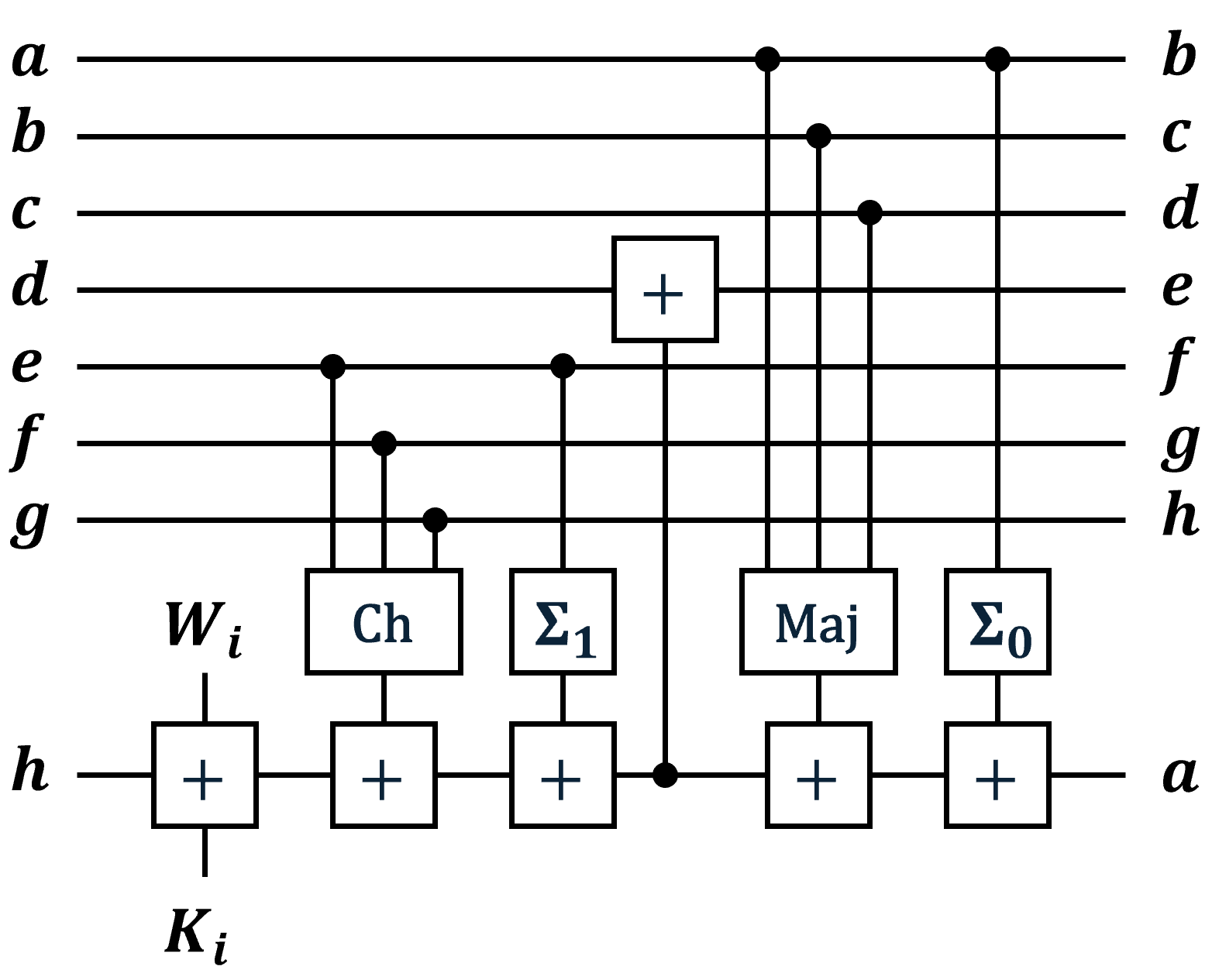}
  \caption{Single SHA-2 Iteration}
  \label{fig:sha2-classical}
\end{figure}

SHA-2 is the latest hash algorithm standard utilizing a Merkle–Damgård construction. The NIST specification lists six functions that vary in strength and output size: SHA-224, SHA-256, SHA-384, SHA-512, SHA-512/224, and SHA-512/256~\cite{FIPS180}. The important difference between these is that SHA-224 and SHA-256 uses 32-bit words and 512-bit chunks, while the others use 64-bit words and 1024-bit chunks. The state consists of eight words, so it's either 256 or 512 bits depending on the word length. The compression function performs 64 iterations on eight working registers as shown in Fig.~\ref{fig:sha2-classical}.

Here, $W_i$ again comes from a message schedule generated by extending the 16-word chunk into 64 words. (The steps to generate the message schedule are slightly more complex than SHA-1, but the effect is similar.) ``Ch'' is short for bitwise choice. In this case, each bit of the output follows $f$ if $e$ is 1, and $g$ if $e$ is 0. ``Maj'' is short for bitwise majority, where each bit of the output follows whatever value occurs most frequently on the inputs. $\Sigma_1$ and $\Sigma_0$ involve XORing together bitwise rotated copies of the input. Note an intermediate value is added to $d$ partway through the iteration.

\begin{figure}[htbp]
  \centering
  \includegraphics[width=\linewidth]{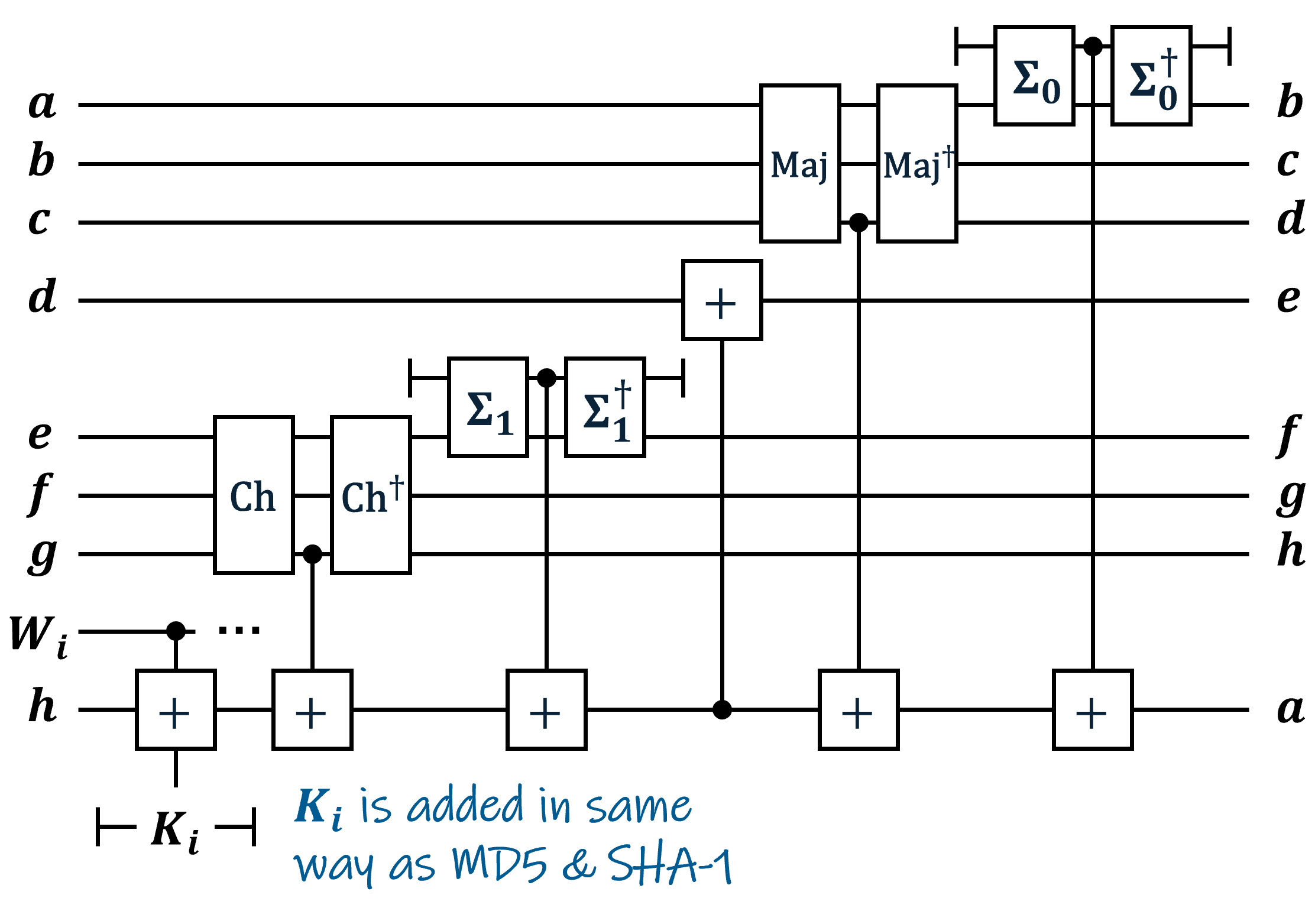}
  \caption{Single SHA-2 Iteration}
  \label{fig:sha2-quantum}
\end{figure}

SHA-2 is a little tricker to implement in quantum software because all these functions must be performed in-place and then reversed, as shown in Fig.~\ref{fig:sha2-quantum}. The main thing to notice in the figure is the use of temporary qubits for the $\Sigma$ functions. This is because they cannot be computed in-place and so a register is needed to store the result. Fortunately, the same qubits can be recycled for constant addition and both $\Sigma$ functions, so only one extra word is needed.

Architecturally, the Merkle–Damgård-based functions lend themselves well to quantum implementation. However, the addition operation prevalent throughout these is relatively expensive in computational cost. (In contrast to a classical computer, there is no dedicated ALU to perform the addition; each plus block represents a string of quantum gates implementing a ripple-carry adder.) The situation is opposite for SHA-3 – the hash function is architecturally complex and difficult to implement, but it does not rely on arithmetic addition.

\subsection{SHA-3}

SHA-3 was designed to be an effective, unique alternative to provide resiliency in the case of a catastrophic vulnerability being discovered in SHA-2. It is based on a cryptographic primitive called Keccak, which utilizes a sponge construction to process the input message and generate the output digest~\cite{bertoni2007sponge}. The high-level steps of Keccak are:

\begin{enumerate}
  \item Pad the input message so its length is divisible by a fixed value $r$ (the bitrate) and the final byte is a specified suffix.
  \item Break the input into chunks of length $r$.
  \item Allocate a state register initialized to all zeros (for SHA-3 this is 1600 bits).
  \item ``Absorb'' phase. For each chunk:
  \begin{enumerate}
    \item XOR the chunk with the first $r$ bits of the state.
    \item Apply a complex function to the state called the block permutation.
  \end{enumerate}
  \item ``Squeeze'' phase. While more output is desired:
  \begin{enumerate}
    \item Append the first $r$ bits of the state to the output.
    \item Apply the block permutation.
  \end{enumerate}
\end{enumerate}

The SHA-3 NIST specification defines drop-in replacements for SHA-2 functions with the same level of security: SHA3-224, SHA3-256, SHA3-384, and SHA3-512~\cite{FIPS202}. It also introduces two functions with variable-length output, SHAKE128 and SHAKE256. These all follow the above steps, but with different values for the padding suffix and bitrate. (Note that a lower bitrate or $r$ value means a higher level of security since it results in more permutations.)

For SHA-3, the devil is in the details of the block permutation. First, the 1600-bit state register is mapped to a 5x5x64 array, and then five functions are applied in succession: $\theta$, $\rho$, $\pi$, $\chi$, and $\iota$. These are composed of only XOR, AND, and NOT gates, plus permutations of the bits in the array, so they are relatively straightforward to implement on a classical computer. Importantly, they are all invertible, i.e., for each function, another function exists that maps the output of the original to the corresponding input, for all such input/output pairs. This is not just theoretical – the Keccak authors provide C++ implementations of the inverse functions in the KeccakTools GitHub repository.\footnote{\url{https://github.com/KeccakTeam/KeccakTools}}

\begin{figure}[htbp]
  \centering
  \includegraphics[width=\linewidth]{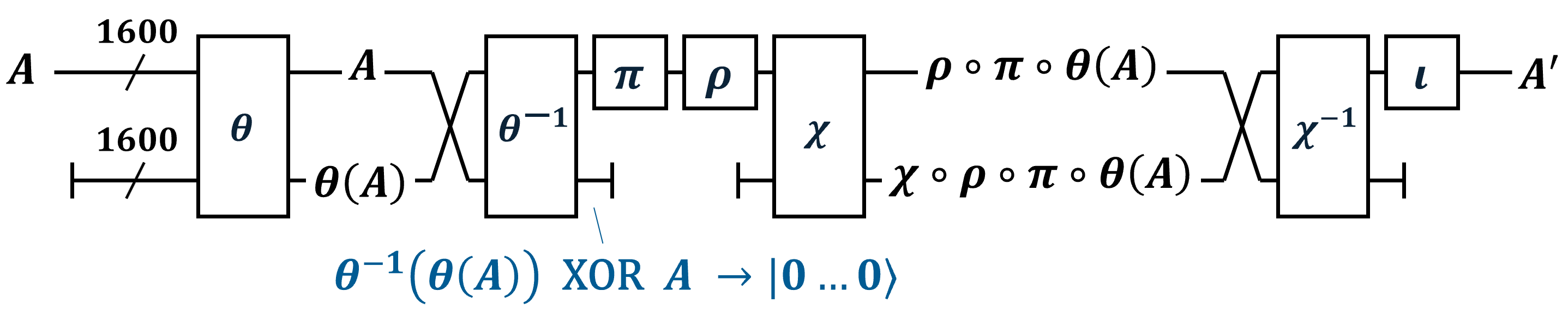}
  \caption{In-Place Keccak Block Permutation Using Inverse Functions}
  \label{fig:keccak-amy}
\end{figure}

Amy et al.\ showed how the block permutation could be performed in-place on a quantum computer~\cite{amy2016estimating}. Since $\rho$ and $\pi$ only consist of permutations or rotations, they could be implemented by reordering the state array (no quantum gates required). And $\iota$ simply XORs the first lane of the state array with a constant that depends on the round, which we have seen before. However, $\theta$ and $\chi$ are trickier; the former involves computing the parities of each column in the state array, and the latter is a non-linear function utilizing AND and NOT. Amy et al.\ mapped the classical steps of these functions directly to quantum gates, so that the output would be constructed in a new 1600-qubit register. Then, they applied the inverse function with the input/output registers swapped, transforming the original input register to all $\ket{0}$’s so those qubits could be released (de-allocated). Fig.~\ref{fig:keccak-amy} illustrates the authors’ strategy for computing the Keccak block permutation in-place in a quantum context.

\begin{figure}[htbp]
  \centering
  \includegraphics[width=\linewidth]{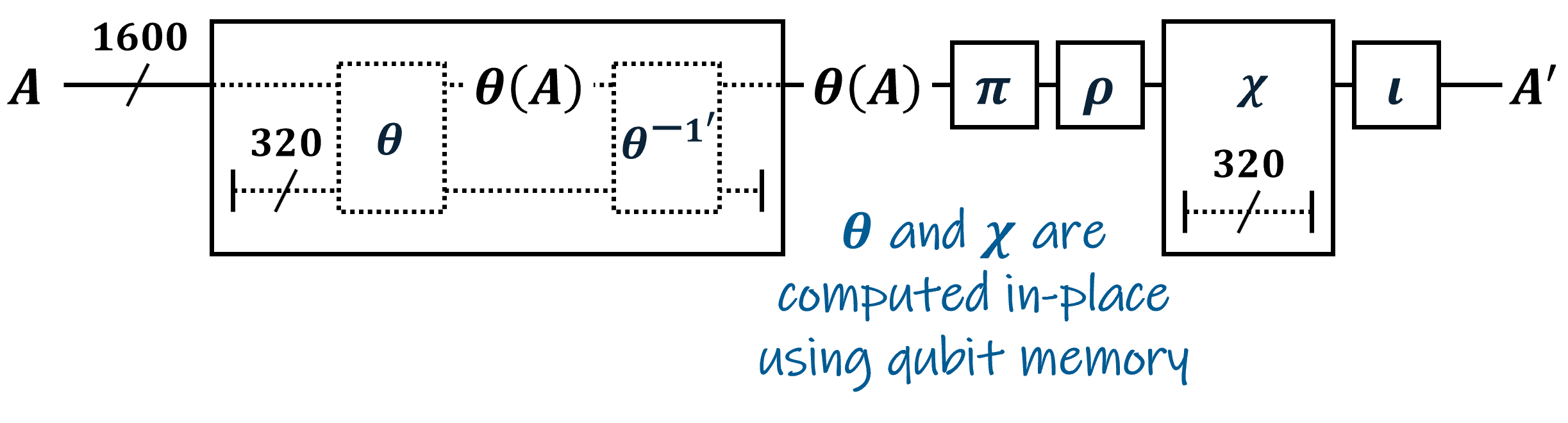}
  \caption{Qubit-Efficient Keccak Block Permutation}
  \label{fig:keccak-preston}
\end{figure}

While the entire block permutation is computed in-place with the above approach, $\theta$ and $\chi$ individually are not. That is, the two functions store the result in an output register, rather than modifying the input register. When implementing these in Q\#, it was discovered that $\theta$ and $\chi$ could themselves be computed in-place, requiring fewer qubits and gates than allocating an output register and computing the inverse. For both functions, a working memory of 320 qubits is used to store intermediate values necessary to transform the input, and then a modified version of the inverse function is used to recompute those intermediate values so the ancillary qubits can be released. This modified inverse is denoted by $\theta^{-1'}$ in Fig~\ref{fig:keccak-preston}; the modified inverse for $\chi$ is not shown. Using the method below, the block permutation can be performed with only 1920 logical qubits, instead of the 3200 required by Amy's approach, without increasing the gate depth.

Listing~\ref{lst:theta} shows the Q\# implementation of the in-place theta function. Note that the modified inverse is built-in so that the ancilla qubits can be scoped to the single operation (they are allocated and de-allocated within it). The code is based on the Keccak team's classical implementation of the inverse theta function,\footnote{\url{https://github.com/KeccakTeam/KeccakTools/blob/master/Sources/Keccak-f.h\#L553}} and uses the following custom operations:

\begin{itemize}
  \item \texttt{ArrayAsWords}: Breaks up a flat array into an array of words with specified length.
  \item \texttt{Xor}: Applies CNOT to each pair of qubits in the parallel array arguments.
  \item \texttt{RightRotate}: Shifts an array to the right in a circular manner by a specified amount. I.e., elements that ``fall off'' the right are ``fed into'' the left.
\end{itemize}

\begin{lstlisting}[caption={In-Place Theta Function}, basicstyle={\ttfamily\scriptsize}, label={lst:theta}]
operation Theta (lanes : Qubit[][][]) : Unit
is Adj + Ctl {
  use temp = Qubit[320];

  // Compute column parities
  let cols = ArrayAsWords(64, temp);
  for x in 0 .. 4 {
    for y in 0 .. 4 {
      Xor(lanes[(x+4) % 5][y], cols[x]);
      Xor(
        RightRotate(lanes[(x+1) % 5][y], 1),
        cols[x]
      );
    }
  }

  // Xor parities into the state array
  for x in 0 ..4 {
    for y in 0 .. 4 {
      Xor(cols[x], lanes[x][y]);
    }
  }

  // Return temp qubits to |0>
  let inversePositions = [
    0xDE26BC4D789AF134L,
    0x09AF135E26BC4D78L,
    0xEBC4D789AF135E26L,
    0x7135E26BC4D789AFL,
    0xCD789AF135E26BC4L
  ];
  for z in 0 .. 63 {
    for xOff in 0 .. 4 {
      if (
        (inversePositions[xOff] >>> z &&& 1L)
        != 0L
      ) {
        for x in 0 .. 4 {
          for y in 0 .. 4 {
            let C = lanes[(x+5-xOff) % 5][y];
            Xor(RightRotate(C, z), cols[x]);
          }
        }
      }
    }
  }
}
\end{lstlisting}

Listing~\ref{lst:chi} shows the Q\# implementation of the in-place chi function. Again, the KeccakTools repository was referenced,\footnote{\url{https://github.com/KeccakTeam/KeccakTools/blob/master/Sources/Keccak-f.h\#L519}} and it uses the following additional custom operations:

\begin{enumerate}
  \item \texttt{Not}: Applies X to each qubit in the array.
  \item \texttt{And}: Applies CCNOT to each triple in the parallel array arguments.
\end{enumerate}

\begin{lstlisting}[caption={In-Place Chi Function}, basicstyle={\ttfamily\scriptsize}, label={lst:chi}]
operation Chi (lanes : Qubit[][][]) : Unit
is Adj + Ctl {
  for y in 0 .. 4 {
    use temp = Qubit[320];

    let rows = ArrayAsWords(64, temp);

    // Copy rows
    for x in 0 .. 4 {
        Xor(lanes[x][y], rows[x]);
    }

    for x in 0 .. 4 {
      // lanes[x][y] = ~rows[x+1] & rows[x+2]
      within {
        Not(rows[(x+1) % 5]);
      }
      apply {
        And(
          rows[(x+1) % 5],
          rows[(x+2) % 5],
          lanes[x][y]
        );
      }
    }

    // Return temp qubits to |0>
    for x in 0 .. 2 .. 6 {
      within {
        Not(lanes[(x+1) % 5][y]);
      }
      apply {
        And(
          lanes[(x+1) % 5][y],
          rows[(x+2) % 5],
          rows[x % 5]
        );
      }
      Xor(lanes[x % 5][y], rows[x % 5]);
    }

    // Compute rows[3] directly
    within {
      Not(lanes[1][y]);
      And(lanes[1][y], lanes[2][y], rows[0]);
      Xor(lanes[0][y], rows[0]);
      Not(lanes[4][y]);
    }
    apply {
      And(lanes[4][y], rows[0], rows[3]);
    }
    Xor(lanes[3][y], rows[3]);
  }
}
\end{lstlisting}

\section{Results} \label{results}

When characterizing a quantum program, the most relevant costs are (1) qubit width, because it must not exceed the number of qubits supported by the hardware, and (2) gate depth, because it corresponds to the amount of time the system state must remain coherent (i.e., no errors) before the result is measured. There are other factors to consider, such as the types and quantities of gates, but from a software perspective, the time complexity of the algorithm is exhibited in the gate depth, and the space complexity is seen in the width. Thus, by varying a parameter and estimating these metrics, we can see its effect on the computational complexity of a quantum program.

\subsection{Computational Complexity of Grover's Algorithm}

Table~\ref{table:search-size} lists the depth and width for different search space sizes when applying Grover’s algorithm to MD5 with a single search target. (Recall the search space is the possible combinations of the input message, and the number of search targets is how many of those combinations are expected to map to a specified output.) Notice that the width remains constant – this is because the input is padded to a multiple of 512 inside the hash function, so the total number of qubits required is the same for the values listed.

\begin{table}[htbp]
  \caption{Cost of MD5 Preimage for Various Search Space Sizes}
  \centering
  \begin{tabular}{ c|c|c }
    \# of Input Qubits& Depth & Width \\
    \hline
    8 & 1,882,132 & 801 \\
    16 & 20,708,556 & 801 \\
    24 & 466,092,196 & 801 \\
    32 & 7,461,175,528 & 801 \\
  \end{tabular}
  \label{table:search-size}
\end{table}

\begin{figure}[htbp]
  \centering
  \includegraphics[width=\linewidth]{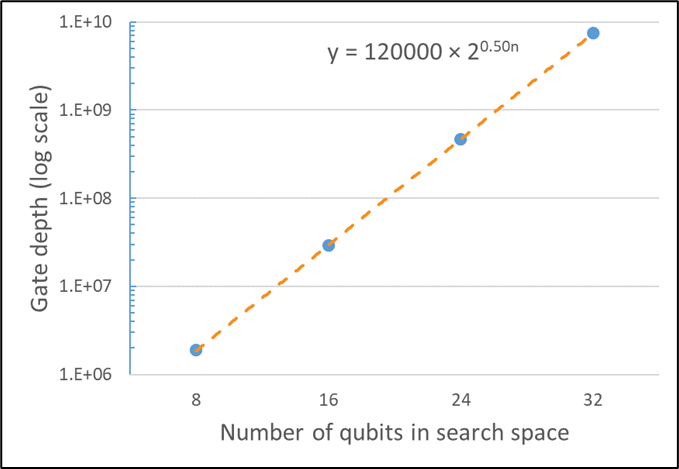}
  \caption{Gate Depth over Search Space Size for MD5 Preimage}
  \label{fig:search-size-plot}
\end{figure}

Fig.~\ref{fig:search-size-plot} is a plot of the data on a log scale. It includes an exponential regression that shows the gate depth increases proportional to $2^{0.5n}$, with $n$ being the number of qubits in the search space. This is consistent with the theoretical time complexity of Grover’s algorithm.\footnote{The size of the search space $N$ is $2^n$, and the number of search targets $k$ is 1, so the required number of Grover's Algorithm iterations is $\mathrm{O}(\sqrt{\frac{N}{k}})=\mathrm{O}(2^{\frac{n}{2}})$.}

\clearpage

\begin{figure}[htbp]
  \centering
  \includegraphics[width=\linewidth]{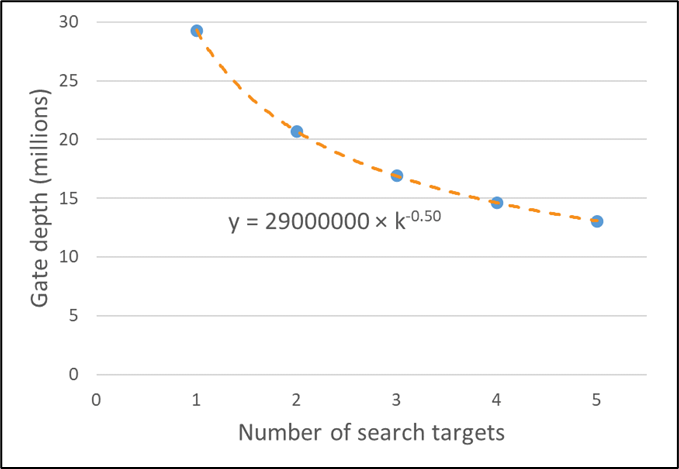}
  \caption{Gate Depth over Search Space Size for MD5 Preimage}
  \label{fig:search-target-plot}
\end{figure}

Table~\ref{table:search-targets} lists the results of a similar analysis, but this time varying the number of search targets instead of the size of the search space. (In this case, the search space was held constant at 16 qubits.) Fig.~\ref{fig:search-target-plot} plots the data with a power regression confirming that the gate depth falls off proportional to $\frac{1}{\sqrt{k}}$, with $k$ being the number of search targets. Again, this matches the expected time complexity.

\begin{table}[htbp]
  \caption{Cost of MD5 Preimage for Various Numbers of Search Targets}
  \centering
  \begin{tabular}{ c|c|c }
    \# of Search Targets & Depth & Width \\
    \hline
    1 & 29,252,464 & 801 \\
    2 & 20,708,556 & 801 \\
    3 & 16,943,444 & 801 \\
    4 & 14,626,452 & 801 \\
    5 & 13,033,520 & 801 \\
  \end{tabular}
  \label{table:search-targets}
\end{table}

While these are not actionable results in and of themselves, they serve as a ``sanity check'' that Grover’s algorithm was implemented correctly, and that the underlying methods used are sound.

\subsection{Differences Between Hash Functions}

The main result of this study is the data in Table~\ref{table:hash-functions}, which lists the required gate depth and qubit width to conduct a preimage attack on each of the hash functions implemented. The number of qubits in the search space was kept constant at 16, with only one search target.

\begin{table}[htbp]
  \caption{Cost of Preimage For Various Hash Functions}
  \centering
  \begin{tabular}{ c|c|c }
    Hash Function & Depth & Width \\
    \hline
    MD5 & 29,252,464 & 801 \\
    SHA-1 & 36,475,332 & 2913 \\
    SHA-224 & 38,953,918 & 2593 \\
    SHA-256 & 38,953,918 & 2593 \\
    SHA-384 & 99,089,086 & 6209 \\
    SHA-512 & 99,089,086 & 6209 \\
    SHA-512/224 & 99,089,086 & 6209 \\
    SHA-512/256 & 99,089,086 & 6209 \\
    SHA3-224 & 1,131,675 & 2161 \\
    SHA3-256 & 1,131,675 & 2193 \\
    SHA3-384 & 1,131,675 & 2193 \\
    SHA3-512 & 1,505,375 & 2449 \\
    SHAKE128-256 & 1,131,675 & 2193 \\
    SHAKE256-512 & 1,505,375 & 2449 \\
  \end{tabular}
  \label{table:hash-functions}
\end{table}

\begin{figure}[htbp]
  \centering
  \includegraphics[width=\linewidth]{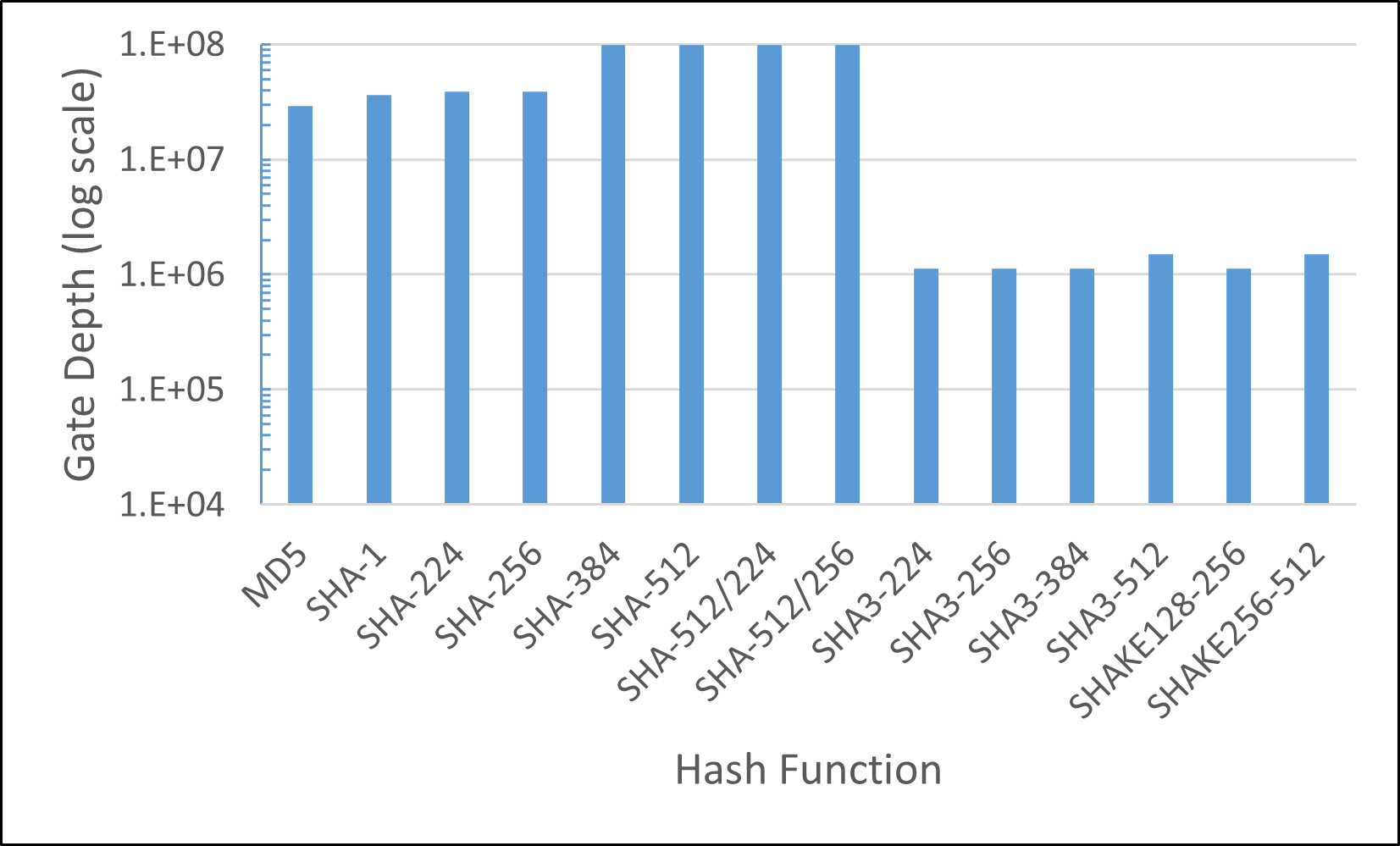}
  \caption{Gate Depth for Preimage of Various Hash Functions}
  \label{fig:gate-depth-plot}
\end{figure}

Fig.~\ref{fig:gate-depth-plot} compares the gate depth of each function. Note the log scale. Clearly, the computational cost of a quantum attack against SHA-3 is much lower than that of MD5, SHA-1, and SHA-2. This is largely due to the fact that SHA-3 utilizes operations and permutations that translate more naturally to a quantum context, while the others rely primarily on arithmetic addition which is expensive on a quantum computer.

\begin{figure}[htbp]
  \centering
  \includegraphics[width=\linewidth]{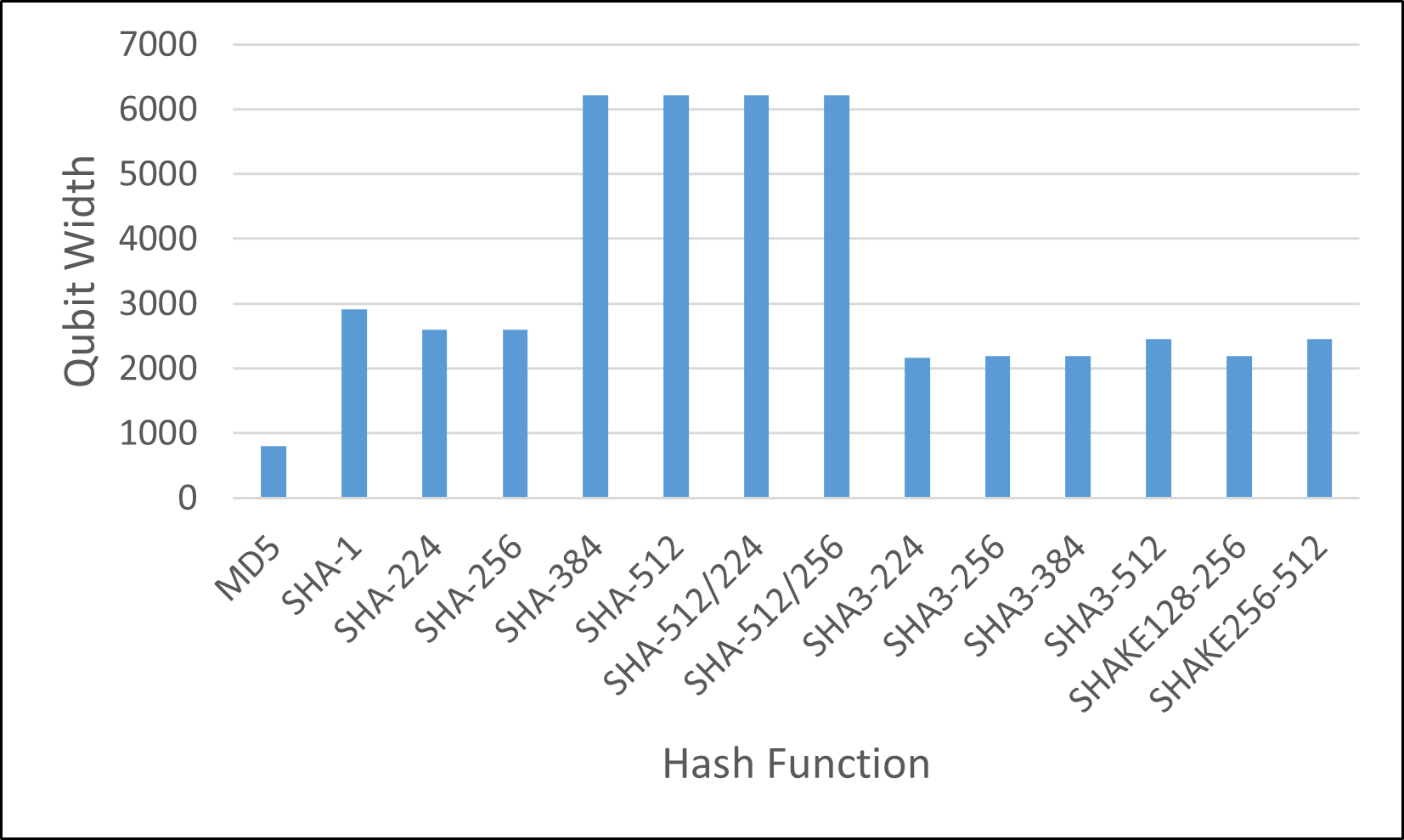}
  \caption{Qubit Width for Preimage of Various Hash Functions}
  \label{fig:qubit-width-plot}
\end{figure}

The picture in Fig.~\ref{fig:qubit-width-plot} is a little different. When it comes to the number of qubits required, SHA-3 is comparable in cost to SHA-2 functions with a 256-bit state, but those with a 512-bit state are much more expensive. This suggests that, for example, SHA-512/256 is more quantum-resistant than SHA-256 and SHA3-256.

\section{Discussion} \label{discussion}

The results obtained show important differences in the computational costs associated with each hash function. Of particular interest is how SHA-3, a more recently developed algorithm, is cheaper to preimage on a quantum computer than the others (expect for MD5 in terms of qubit width). However, it should be noted that the quantum implementations of the hash functions used here are not guaranteed to be optimal. We directly translated the classical steps into quantum instructions, but new methods for creating more cost-efficient quantum oracles could be devised in the future.

The extent of the findings is also qualified by the limitations in the underlying quantum computing hardware, as well as some practical considerations when it comes to conducting preimage attacks against hash functions. These are discussed below.

\subsection {Hardware Limitations}

This paper takes a relatively abstract view of quantum computing, essentially at the level of quantum software. While Q\# provides a simulator and resource estimator, support for running on a real live quantum computer is limited. And so, we do not consider computational costs associated with the underlying quantum hardware. The most important of these is quantum error correction.

On real quantum computers, the qubits are subjected to environmental noise that degrades or destroys their state over time. To perform useful computation, any errors that were introduced by noise must be fixed using quantum error correction (QEC). In a nutshell, QEC uses error codes to reduce the likelihood of error by introducing additional qubits. For example, a QEC scheme might use eight physical qubits to compose a single “logical” qubit, whose error rate is brought down to an acceptable level. Obviously, performing the error correction algorithm introduces additional costs in terms of both qubits and gates. These costs are so prohibitive to current-generation quantum computers that they are hard-pressed to maintain even a single logical qubit for practical applications.\footnote{See Section 3.2 of the National Academies of Sciences, Engineering, and Medicine report, ``Quantum Computing: Progress and Prospects'' for an excellent discussion of how the overhead associated with QEC is a major barrier to realizing scalable quantum computers~\cite{NAP25196}.} This means that any quantum program relying on error-free qubits (such as the ones discussed in this report), cannot be run at all in the near term, and will have large overhead costs associated with QEC into the future.

Another important hardware-related cost is introduced by the qubit topology. There are multiple approaches to how a qubit is physically realized in quantum hardware, and how they are arranged in space. For example, the qubits may be laid out in a 2D grid, and controlled gates such as CNOT can only be applied to adjacent qubits~\cite{kitaev2003}. In that case, SWAP gates would be required to position the desired states in the appropriate location before each CNOT. To put it plainly, this becomes a total mess – you need optimization algorithms to minimize the number of SWAPs, and it may significantly increase the level of computational complexity. It remains to be seen what quantum hardware approach will be most successful, so for now it suffices to mention that this is likely to be a factor in any physical implementation of a quantum algorithm.

\subsection {Practical Considerations of Preimage Attacks}

When talking about a preimage attack against a cryptographic hash algorithm, typically no knowledge of the original input is assumed. That is, you are given a hash output and asked to find at least one corresponding input of any length. Assuming a theoretically perfect hash function, given any distinct input, there is an equal chance ($\frac{1}{2^l}$) of getting any of the possible $l$-bit outputs. Thus, the probability of getting a specific output after $n$ guesses follows a binomial distribution, $P(n)=1-(1-\frac{1}{2}^l)^n$. Setting $n=2^l$ gives a probability of
$$
P(2^l)=1-(1-\frac{1}{2^l})^{2^l}\approx 1-\frac{1}{e} \textrm{ for large } l
$$
That is, the probability that a preimage attack succeeds with $2^l$ guesses and no prior knowledge is about 63.2\%. Of course, in a classical context, if you do not find a match you can just keep guessing and expect to find one eventually.

Now, consider Grover’s algorithm. It is not guess-and-check. Rather, it finds the matching input directly using amplitude amplification. However, the number of search targets must be specified. This is a problem because hash functions are not one-to-one; more than one distinct input can map to the same output.\footnote{Multiple inputs that produce the same output of a hash function are called collisions. Part of the security model of hash functions is that collisions, though they may exist, are practically impossible to find. One way to degrade the security of a hash function is through a birthday attack, which attempts to generate a collision. The BHT algorithm is a quantum birthday attack that makes use of Grover’s Algorithm to generate a collision in $\mathrm{O}(2^{\frac{n}{3}})$ steps, with $n$ being the number of output bits~\cite{Brassard_1998}.} In other words, the number of search targets is not known in advance (although it follows a binomial probability distribution).

The software program we developed for this research supports simple strategies for trying Grover’s algorithm with different numbers of search targets. However, they were not used when estimating computational resources; one search target was assumed. This is because the search space was so small ($2^{16}$) that the probability of a collision existing is near zero. (There are more sophisticated methods of dealing with an unknown number of search targets; Boyer et al.\ lay out one such strategy in ``Tight bounds on quantum searching''~\cite{Boyer_1998}.)

In practice, and particularly for the current officially approved functions (SHA-2 and SHA-3), a search space of sufficient size to contain a collision with non-negligible probability is too big to search, even for Grover’s algorithm. Consider SHA-256. The search space would have to be somewhere near 256 bits before it is likely to contain a collision. But the gate depth of Grover’s algorithm applied to SHA-256 is $\mathrm{O}(2^{\frac{l}{2}})$, or $\mathrm{O}(2^{128})$. Even supposing one trillion operations per second, this would still take around $10^{19}$ years.

A more realistic situation is where the input length is known in advance, or it is at least limited in length. Suppose you intercept the SHA-256 hash of a password that you happen to know is 12 characters long (96 bits). Recovering the password is $\mathrm{O}(2^{96})$ using guess-and-check and $\mathrm{O}(2^{48})$ with Grover’s algorithm on a quantum computer. Assuming one billion operations per second, that is the difference between quadrillions of years and a few days. In scenarios like this, you are searching for exactly one target, and so the results shown in Section~\ref{results} are applicable.

\section {Conclusion} \label{conclusion}

This work provides an example of analyzing quantum algorithms through software. We used Microsoft's QDK to study the practicality of quantum preimage attacks on classical hash functions. This involved implementing the hash functions as quantum oracles in Q\#, validating their correctness using the Toffoli simulator, and then applying the built-in resource estimator to obtain computational complexity metrics. Through this process, we reduced the number of logical qubits required for the SHA-3 oracle by 40\% compared to Amy et al.'s implementation. And our analysis concluded that the 512-bit state variants of SHA-2 are the most quantum-resistant out of the hash functions studied.

The advantage of this approach is that the algorithm must be explicitly expressed as a software program, so it leaves very little room for false assumptions about its actual behavior and complexity. Also, once the program is written and validated, the resource estimation can be performed automatically, rather than manually on paper. As quantum software frameworks like the QDK improve, the metrics reported can be more precise and tailored to the specific hardware platform the program is expected to run on. For example, IBM's Qiskit platform allows the user to run quantum code on real quantum computers or perform simulations that introduce realistic errors~\cite{Qiskit}.

\subsection {Future Work}

Here, we were limited to providing grounded estimates in terms of logical qubits with the QDK, since the width and depth of the programs written for this research were much too large to run on today's NISQ-era devices. It must be emphasized that major blocking challenges exist in creating scalable, error-corrected quantum computers. Nevertheless, this type of analysis is critical in understanding the practical considerations when applying a quantum algorithm to a real-world problem, so that we can prepare appropriately for the threats and opportunities that advances in quantum computing technology will bring.

The logical next step would be to expand the library of classical functions implemented in quantum software so that Grover's Algorithm can be applied. For example, Jaques et al.\ provide oracle implementations for AES-128, -192, and -256 in their paper, ``Implementing Grover Oracles for Quantum Key Search on AES and LowMC''~\cite{eurocrypt-2020-30189}. We could also improve the confidence and precision of our study by reproducing the implementations in other quantum software frameworks, such as Qiskit, and comparing the results of various resource estimation tools.

Going further, the approach we took can be replicated for other quantum algorithms. A standard library of common quantum applications implemented with multiple software frameworks could provide clear and precise benchmarks for quantum computers. The Standards and Performance Metrics Technical Advisory Committee of the Quantum Economic Development Consortium recently published a study in this vein where they developed and proposed application-oriented performance benchmarks and open-sourced their code\footnote{\url{https://github.com/SRI-International/QC-App-Oriented-Benchmarks}}~\cite{lubinski2021applicationoriented}. By providing actual quantum programs to run, different quantum computers can be compared fairly, and the practicality of running an algorithm on various quantum systems can be studied in detail.

\bibliographystyle{unsrtnat}
\bibliography{preston}

\end{document}